\documentclass[sigplan,10pt,review]{acmart}

\acmConference[PL'17]{ACM SIGPLAN Conference on Programming Languages}{January 01--03, 2017}{New York, NY, USA}
\acmYear{2017}
\acmISBN{} 
\acmDOI{} 
\startPage{1}

\setcopyright{none}

\usepackage{booktabs}   
\usepackage{subcaption} 
\usepackage{graphicx}
\usepackage{adjustbox}
\usepackage{xspace}
\usepackage{enumitem}
\usepackage{amssymb}
\usepackage{amsmath}
\usepackage{amsthm}
\usepackage[linesnumbered,ruled,vlined]{algorithm2e}
\usepackage{mathtools,cases}
\usepackage{dsfont}
\usepackage{scalerel,stackengine}
\usepackage{listings}
\usepackage{pifont}
\usepackage{xcolor}
\usepackage{soul}
\usepackage{upgreek}

\usepackage{wrapfig}

\usepackage[scaled]{beramono}
\usepackage[T1]{fontenc}

\usepackage{microtype}

\usepackage{tikz}

\usepackage{balance}

\usepackage{fancyhdr}
\theoremstyle{definition}
\newtheorem{definition}{Definition}[section]

\newcommand{\cmark}{\ding{51}}%
\newcommand{\xmark}{\ding{55}}%

\newcommand{\etal}{\mbox{\em et al.}\xspace}
\newcommand{\ie}{\mbox{\em i.e.}\xspace}
\newcommand{\eg}{\mbox{\em e.g.}\xspace}
\newcommand{\wrt}{\mbox{\em w.r.t.}\xspace}
\newcommand{\cf}{\mbox{\em cf.}\xspace}
\newcommand{\etc}{\mbox{\em etc.}\xspace}

\newcommand{\course}{\mbox{Microsoft-DEV204.1x}\xspace}
\newcommand{\tool}{\textsc{Sarfgen}\xspace}
\newcommand{\pavector}{\mbox{position-aware characteristic vector}\xspace}

\newcommand{\Csharp}{{\settoheight{\dimen0}{C}C\kern-.05em \resizebox{!}{\dimen0}{\raisebox{\depth}{\#}}}\xspace}

\makeatletter
\newcommand*{\rom}[1]{\expandafter\@slowromancap\romannumeral #1@}
\makeatother

\def\apeqA{\SavedStyle\sim}
\def\apeq{\setstackgap{L}{\dimexpr.5pt+1.5\LMpt}\ensurestackMath{%
		\ThisStyle{\mathrel{\Centerstack{{\apeqA} {\apeqA} {\apeqA}}}}}}

\DeclareMathOperator*{\argminA}{arg\,min}
\newcommand*{\Scale}[2][4]{\scalebox{#1}{$#2$}}%

\begin{document}

\title[Data-Driven Feedback Generation for Introductory Programming Exercises]{Search, Align, and Repair: Data-Driven Feedback Generation for Introductory Programming Exercises}           


\author{Ke Wang}
\affiliation{
	\institution{University of California, Davis}            
}
\email{kbwang@ucdavis.edu}          

\author{Rishabh Singh}
\affiliation{
	\institution{Microsoft Research}            
}
\email{risin@microsoft.com} 

\author{Zhendong Su}
\affiliation{
	\institution{University of California, Davis}            
}
\email{su@cs.ucdavis.edu} 


\begin{abstract}

This paper introduces the ``Search, Align, and Repair'' data-driven
program repair framework to automate feedback generation for introductory programming exercises.  Distinct from existing techniques, our goal is to
develop an efficient, fully automated, and problem-agnostic technique for large or MOOC-scale introductory
programming courses. We leverage the large amount of available student
submissions in such settings and develop new algorithms for
identifying similar programs, aligning correct and incorrect programs,
and repairing incorrect programs by finding minimal fixes. We have
implemented our technique in the \tool system and evaluated it on
thousands of real student attempts from the \course edX
course and the Microsoft CodeHunt platform. Our results show that \tool
can, within two seconds on average, generate concise, useful feedback for 89.7\%
of the incorrect student submissions. It has been integrated with the Microsoft-DEV204.1X edX class and
deployed for production use.

\end{abstract}

%


\maketitle

\makeatletter
\newenvironment{btHighlight}[1][]
{\begingroup\tikzset{bt@Highlight@par/.style={#1}}\begin{lrbox}{\@tempboxa}}
	{\end{lrbox}\bt@HL@box[bt@Highlight@par]{\@tempboxa}\endgroup}

\newcommand\btHL[1][]{%
	\begin{btHighlight}[#1]\bgroup\aftergroup\bt@HL@endenv%
	}
	\def\bt@HL@endenv{%
	\end{btHighlight}%
	\egroup
}
\newcommand{\bt@HL@box}[2][]{%
	\tikz[#1]{%
		\pgfpathrectangle{\pgfpoint{1pt}{0pt}}{\pgfpoint{\wd #2}{\ht #2}}%
		\pgfusepath{use as bounding box}%
		\node[anchor=base west, fill=orange!30,outer sep=0pt,inner xsep=.5pt, inner ysep=-0.3pt, rounded corners=2pt, minimum height=\ht\strutbox+1pt,#1]{\raisebox{.01pt}{\strut}\strut\usebox{#2}};
	}%
}
\makeatother

\definecolor{codegreen}{rgb}{0,0.6,0}
\definecolor{codegray}{rgb}{0.5,0.5,0.5}
\definecolor{codepurple}{rgb}{0.58,0,0.82}

\lstdefinestyle{mystyle}{
	commentstyle=\color{codegreen},
	keywordstyle=\color{blue}\bfseries,
	numberstyle=\tiny\color{codegray},
	stringstyle=\color{codepurple},
	basicstyle=\linespread{0.85}\fontsize{5.5}{8.8}\ttfamily,
	breakatwhitespace=false,         
	breaklines=true,                 
	captionpos=b,                    
	keepspaces=true,                 
	numbers=left,                    
	numbersep=3pt,                  
	showspaces=false,                
	showstringspaces=false,
	showtabs=false,                  
	tabsize=2,
	language=[Sharp]C,
	moredelim=**[is][\btHL]{`}{`},
	moredelim=**[is][{\btHL[fill=red!40]}]{@}{@},
}

\setul{}{1.2pt}

\section{Introduction}
\label{section:intr}

The unprecedented growth of technology and computing related jobs in
recent years~\cite{soper2014analysis} has resulted in a surge in
Computer Science enrollments at colleges and universities, and
hundreds of thousands of learners worldwide signing up for Massive
Open Online Courses (MOOCs). While larger classrooms and MOOCs have
made education more accessible to a much more diverse and greater
audience, there are several key challenges that remain to ensure
comparable education quality to that of traditional smaller classroom
settings. This paper tackles one such challenge of \emph{providing
  fully automated, personalized feedback to students for introductory
  programming exercises without requiring any instructor effort}.

The problem of automated feedback generation for introductory
programming courses has seen much recent interest --- many systems
have been developed using techniques from formal methods, programming
languages, and machine learning. Most of these
techniques model the problem as a \emph{program repair} problem:
repairing an incorrect student submission to make it functionally
equivalent \wrt a given specification (\eg, a reference solution or a
set of test cases). Table~\ref{Table:comp} summarizes some of the 
major techniques in terms of their capabilities and
requirements, and compares them with our proposed technique realized
in the \tool\footnote{\underline{S}earch, \underline{A}lign, and
  \underline{R}epair for \underline{F}eedback \underline{GEN}eration}
system.


As summarized in the table, existing systems still face important
challenges to be effective and practical. In particular,
AutoGrader~\cite{singh2013automated} requires a custom error model per
programming problem, which demands manual effort from the instructor
and her understanding of the system details. Moreover, its reliance on
constraint solving to find repairs makes it expensive and unsuitable in
interactive settings at the MOOC scale. Systems like
CLARA~\cite{gulwani2016automated} and sk\_p~\cite{pu2016sk_p} can
generate repairs relatively quickly by using clustering and machine
learning techniques on student data, but the generated repairs are
often imprecise and not minimal, which do not provide quality
feedback. CLARA's use of Integer Linear Programming (ILP) for variable
alignment during repair generation also hinders its
scalability. Section~\ref{sec:rel} provides a detailed survey of
related work.

\begin{table}[tbp!]
	\begin{center}
		\begin{adjustbox}{max width=.46\textwidth}
			\begin{tabular}{|c | c | c | c | c | c | c | c} 
				\hline
				
				\textbf{Approach} 
				& \begin{tabular}{@{}c@{}}\textbf{No Manual} \\ \textbf{Effort}\end{tabular}
				& \begin{tabular}{@{}c@{}}\textbf{Minimal} \\ \textbf{Repair}\end{tabular}
				& \begin{tabular}{@{}c@{}}\textbf{Complex} \\ \textbf{Repairs}\end{tabular}
				& \textbf{Data-Driven}
               & \begin{tabular}{@{}c@{}}\textbf{Production} \\ \textbf{Deployment}\end{tabular}\\ 

				\hline \hline
				AutoGrader~\cite{singh2013automated} &\xmark  &\cmark  &\xmark &\xmark &\xmark \\						
				\hline
				CLARA~\cite{gulwani2016automated} &\cmark &\xmark &\cmark &\cmark  &\xmark \\						
				\hline
				QLOSE~\cite{d2016qlose} &\xmark &\xmark &\xmark &\xmark  &\xmark \\						
				\hline
				sk\_p~\cite{pu2016sk_p} &\cmark &\xmark &\cmark &\cmark  &\xmark \\						
				\hline
				REFAZER~\cite{rolim2017learning} &\cmark &\xmark &\xmark &\cmark  &\xmark \\						
				\hline
				CoderAssist~\cite{Kaleeswaran} &\xmark &\xmark &\cmark &\cmark  &\xmark \\						
				\hline
				\hline
				\textbf{\tool} &\cmark &\cmark &\cmark &\cmark  &\cmark \\						
				\hline
				
			\end{tabular}
		\end{adjustbox}
	\end{center}
	\caption{Comparison of \tool against the existing feedback generation approaches.}
	\vspace{-.5cm}
	\label{Table:comp}
\end{table}

To tackle the weaknesses of existing systems, we introduce
``Search, Align, and Repair'', a conceptual framework for program 
repair from a data-driven perspective. First, given an incorrect program, 
we search for similar reference solutions. Second, we align 
each statement in the incorrect program with a corresponding statement in the 
reference solutions to identify discrepancies for 
suggesting changes. Finally, we pinpoint minimal fixes to patch 
the incorrect program. Our key observation is that the diverse
approaches and errors students make are captured in the abundant
previous submissions because of the MOOC
scale~\cite{drummond2014learning}. Thus, we aim at a fully automated, data-driven
approach to generate \emph{instant} (to be interactive), \emph{minimal} (to be
precise) and \emph{semantic} (to allow complex repairs) fixes to
incorrect student submissions. At the technical level, we need to address three
key challenges:
\begin{description}
\item[Search:] Given an incorrect student program, how to efficiently
  identify a set of closely-related candidate programs among all
  correct solutions?

\item[Align:] How to efficiently and precisely align each selected program with the incorrect
  program to compute a correction set that consists of expression- or
  statement-level discrepancies?

\item[Repair:] How to quickly identify a minimal set of fixes out
  of an entire correction set? 
\end{description}

For the ``Search'' challenge, we identify \textit{syntactically} most
similar correct programs \wrt the incorrect program in a
coarse-to-fine manner. First, we perform exact matching on the control
flow structures of the incorrect and correct programs, and rank
matched programs \wrt the tree edit distance between their abstract
syntax trees (ASTs). Although introductory programming assignments
often feature only lightweight small programs, the massive number of
submissions in MOOCs still makes the standard tree edit distance
computation too expensive for the setting. Our solution is based on a
new tree embedding scheme for programs using numerical embedding
vectors with a new distance metric in the Euclidean space. The new program embedding vectors (called \emph{position-aware characteristic vectors}) efficiently capture the structural information of the program ASTs. Because the
numerical distance metric is easier and faster to compute,
program embedding allows us to scale to a large number of programs.

For the ``Align'' challenge, we propose a usage-based 
$\alpha$-conversion to rename variables in a correct program using 
those from the incorrect program. In particular, we represent each 
variable with the new embedding vector based on its usage profile 
in the program, and compute the mapping between two sets of 
variables using the Euclidean distance metric. In the next phase, 
we split both programs into sequences of basic blocks and align 
each pair accordingly. Finally, we construct discrepancies by 
matching statements only from the aligned basic blocks.


For the final ``Repair'' challenge, we \textit{dynamically} minimize
the set of corrections needed to repair the incorrect program from the
large set of possible corrections generated by the alignment step. We 
present some optimizations that gain significant speed-up 
over the enumerative search.



Our automated program repair technique offers several important benefits:
\begin{itemize}
  \item \textbf{Fully Automated}: It does not require any manual
    effort during the complete repair process.

  \item \textbf{Minimal Repair}: It produces a \textit{minimal} set of
    corrections (\ie, any included correction is relevant and
    necessary) that can better help students.

  \item \textbf{Unrestricted Repair}: It supports both simple and
    complex repair modifications to the incorrect program.

  \item \textbf{Portable}: Unlike most previous approaches, it is
    independent of the programming exercise ---
    it only needs a set of correct submissions for each exercise.
\end{itemize}

We have implemented our technique in the $\tool$ system and extensively
evaluated it on thousands of programming submissions obtained from the
\course edX course and the CodeHunt
platform~\cite{codehunt}. $\tool$ is able to repair \textit{89.7}\% of
the incorrect programs with \textit{minimal} fixes in under two seconds on
average. In addition, it has been integrated with the
\course course and deployed for production use. The feedback collected from online students demonstrates its practicality and usefulness.

This paper makes the following main contributions: 
\begin{itemize}
  \item We propose a high-level data-driven framework --- search, align and repair --- 
  to fix programming submissions for introductory programming exercises. 
  
  \item We present novel instantiations for different framework components. Specifically, we develop novel program embeddings and the 
  associated distance metric to efficiently and precisely identify 
  similar programs and compute program alignment.
  

  \item We present an extensive evaluation of $\tool$ on repairing
    thousands of student submissions on 17 different programming
    exercises from the Microsoft-DEV204- .1x edx course and the Microsoft CodeHunt platform.
\end{itemize}

\section{Overview}\label{section:overiew}

This section gives an overview of our approach by introducing its key
ideas and high-level steps.

\begin{figure}[h]
	\centering
	\texttt{X} \texttt{O} \texttt{X} \texttt{O} \texttt{X} \texttt{O} \texttt{X} \texttt{O} \\
	\texttt{O} \texttt{X} \texttt{O} \texttt{X} \texttt{O} \texttt{X} \texttt{O} \texttt{X} \\
	\texttt{X} \texttt{O} \texttt{X} \texttt{O} \texttt{X} \texttt{O} \texttt{X} \texttt{O} \\
	\texttt{O} \texttt{X} \texttt{O} \texttt{X} \texttt{O} \texttt{X} \texttt{O} \texttt{X} \\
	\texttt{X} \texttt{O} \texttt{X} \texttt{O} \texttt{X} \texttt{O} \texttt{X} \texttt{O} \\
	\texttt{O} \texttt{X} \texttt{O} \texttt{X} \texttt{O} \texttt{X} \texttt{O} \texttt{X} \\
	\texttt{X} \texttt{O} \texttt{X} \texttt{O} \texttt{X} \texttt{O} \texttt{X} \texttt{O} \\
	\texttt{O} \texttt{X} \texttt{O} \texttt{X} \texttt{O} \texttt{X} \texttt{O} \texttt{X} \\	
	\caption{Desired output for the chessboard exercise.}
	\label{fig:res}
\end{figure}

\subsection{Example: The Chessboard printing problem}

The Chessboard printing assignment, taken from the edX course of \Csharp
programming, requires students to print the pattern of chessboard using
"\texttt{X}" and "\texttt{O}" characters to represent the squares as shown in
Figure~\ref{fig:res}.

%

\begin{figure*}[t]
	\begin{subfigure}[b]{0.32\textwidth}
		\lstset{style=mystyle}
\begin{lstlisting}[escapechar=@,basicstyle=\fontsize{7}{11.2}\ttfamily]
static void Main(string[] args){
    for(int i = 0;i < 8;i++){
        String chess = @""@;
        for (int j = 0;j < 8;j++){
            if (j % 2 == 0)
                chess += "X";
            else
                chess += "0";
        }
        Console.Write(chess+"\n");
    }}

\end{lstlisting}		
		\caption{}
		\label{fig:ic1}
	\end{subfigure}
	\; 
	\begin{subfigure}[b]{0.32\textwidth}		
		\lstset{style=mystyle}
\begin{lstlisting}[escapechar=@,basicstyle=\linespread{.65}\fontsize{7}{11.2}\ttfamily]
static void Main(string[] args){
    int length = 8;
    bool ch = true;

    for (int i = 0; i < length; i++)
    {
        for (int j = 0; j < length; j++)
        {
            if (ch) Console.Write("X");
            else Console.Write("O");

            ch = !ch;
        }

        Console.Write("\n");
    }}
\end{lstlisting}	
		\caption{}
		\label{fig:ic2}
	\end{subfigure}
	\; 
	\begin{subfigure}[b]{0.32\textwidth}
		\lstset{style=mystyle}
\begin{lstlisting}[escapechar=@,basicstyle=\linespread{.65}\fontsize{7}{11.2}\ttfamily]
static void Main(string[] args){
    for (int i = @1@; i < @9@; i++){
        for (int j = @1@; j < @9@; j++){
            if (i % 2 > 0)
                if (j % 2 @>@ 0)
                    Console.Write(@"O"@);
                else
                    Console.Write(@"X"@);
            else
                if (j % 2 @>@ 0)
                    Console.Write(@"X"@);
                else
                    Console.Write(@"O"@);
        }
        Console.WriteLine();
    }}

\end{lstlisting}	
		\caption{}
		\label{fig:ic3}
	\end{subfigure}
	\caption{Three different incorrect student implementations for the chessboard problem.}
	\label{fig:subs}
\end{figure*}

\begin{figure*}[h]
	\begin{subfigure}[b]{0.32\textwidth}
		\lstset{style=mystyle}
\begin{lstlisting}[escapechar=@,basicstyle=\fontsize{7}{11.2}\ttfamily]
static void Main(string[] args){
    for (int i = 0; i < 8; i++){
       String query = @string.Empty@;
       for (int j = 0; j < 8; j++){
            if ((i + j) % 2 == 0)
                query += "X";
            else
                query += "O";
       }
       @Console.WriteLine(query);@
    }}

\end{lstlisting}		
		\caption{}
		\label{fig:c1}
	\end{subfigure}.
	\; 
	\begin{subfigure}[b]{0.32\textwidth}
		\lstset{style=mystyle}
\begin{lstlisting}[escapechar=@,basicstyle=\linespread{.65}\fontsize{7}{11.2}\ttfamily]
static void Main(string[] args){
    bool blackWhite = true;
    for (int i = 0; i < 8; i++)
    {
        for (int j = 0; j < 8; j++)
        {
            if (blackWhite)
                Console.Write("X");
            else
                Console.Write("O");
            blackWhite = !blackWhite;
        }

        blackWhite = !blackWhite;
        Console.WriteLine("");
    }}
\end{lstlisting}		    
		\caption{}
		\label{fig:c2}
	\end{subfigure}
	\; 
	\begin{subfigure}[b]{0.32\textwidth}
		\lstset{style=mystyle}
\begin{lstlisting}[escapechar=@,basicstyle=\linespread{.65}\fontsize{7}{11.2}\ttfamily]
static void Main(string[] args){
    for (int i = @0@; i < @8@; i++){
        for (int j = @0@; j < @8@; j++){
            if (i % 2 == 0)
                if (j % 2 @==@ 0)
                    Console.Write(@"X"@);
                else
                    Console.Write(@"O"@);
            else
                if (j % 2 @==@ 0)
                    Console.Write(@"O"@);
                else
                    Console.Write(@"X"@);
        }
        Console.WriteLine();
    }}

\end{lstlisting}	
		\caption{}
		\label{fig:c3}
	\end{subfigure}
	\caption{The respective top-1 reference solutions found by $\tool$.}
	\label{fig:sols}
\end{figure*}

\begin{figure*}[h]
	\begin{subfigure}[b]{0.331\textwidth}
		The program requires \textbf{2} changes:
		\begin{itemize}[leftmargin=10pt]
			\item In \textbf{j \% 2 == 0} on \textbf{line 5}, change \textbf{j} to \textbf{(i+j)}.
			\item In \textbf{chess += 0} on \textbf{line 8}, replace \textbf{0} to \textbf{O}.		
		\end{itemize}
		\vspace{.7pt}
		\caption{}		
	\end{subfigure}
	\quad\;
	\begin{subfigure}[b]{0.28\textwidth}
		The program requires \textbf{1} change:
		\begin{itemize}[leftmargin=10pt]
			\item At \textbf{line 14}, add \textbf{ch = !ch}.
		\end{itemize}
		\vspace{2pt}
		\caption{}		
	\end{subfigure}
	\; 
	\begin{subfigure}[b]{0.321\textwidth}
		The program requires \textbf{1} change:
		\begin{itemize}[leftmargin=10pt]
			\item In \textbf{i \% 2 > 0} on \textbf{line 4}, change \textbf{>} to \textbf{==}.
		\end{itemize}
		\caption{}		
	\end{subfigure}
%
	\caption{The feedback generated by $\tool$ on the three incorrect student submissions from Figure~\ref{fig:subs}.}
	\label{fig:feed}
\end{figure*}

The goal of this problem is to teach students the concept of
conditional and looping constructs. On this problem, students struggled
with many issues, such as loop bounds or conditional predicates.  In
addition, they also had trouble understanding the functional behavior
obtained from combining looping and conditional constructs.  For example,
one common error we observed among student attempts was that they
managed to alternate "\texttt{X}" and "\texttt{O}" for each separate row/column, but
(mistakenly) repeated the same pattern for all rows/columns (rather
than flipping for consecutive rows/columns).

One key challenge in providing feedback on programming submissions is
that a given problem can be solved using many different algorithms.
In fact, among all the correct student submissions, we found 337
different control-flow structures, indicating the fairly large
solution space one needs to consider. It is worth mentioning that
modifying incorrect programs to merely comply with the specification
using a reference solution may be inadequate. For example,
completely rewriting an incorrect program into a correct program will
achieve this goal, however such a repair might lead a student to a
completely different direction, and would not help her understand
the problems with her own approach. Therefore, it is important to
pinpoint minimal modifications in their particular implementation
that addresses the root cause of the problem. In addition, efficiency
is important especially in an interactive setting
like MOOC where students expect instant feedback within few seconds
and also regarding the deployment cost for thousands of students.

Given the three different incorrect student submissions shown in
Figure~\ref{fig:subs}, $\tool$ generates the feedback depicted in
Figure~\ref{fig:feed} in under two seconds each. During these two
seconds, $\tool$ searches over more than two thousand reference
solutions, and selects the programs in Figure~\ref{fig:sols} for each
incorrect submission to be compared against. For brevity, we only show
the top-1 candidate program in the comparison set. $\tool$ then collects and minimizes the
set of changes required to eliminate the errors.  Finally, it produces
the feedback which consists of the following information (highlighted
in bold in Figure~\ref{fig:feed} for emphasis):

\begin{itemize}
	\item The number of changes required to correct the errors in the program.
	
	\item The location where each error occurred denoted by the line number. 

	\item The problematic expression/statement in the line.
	
	\item The problematic sub-expression/expression that needs to be corrected. 
	
	\item The new value of sub-expression/expression. 
\end{itemize} 

$\tool$ is customizable in terms of the level of the feedback an
instructor would like to provide to the students. The generated
feedback could consist of different combinations of the five kinds of
information, which enables a more personalized, adaptive, and dynamic
tutoring workflow.

\subsection{Overview of the Workflow}

\begin{figure}[ht!]
	\centering
	\includegraphics[width=0.48\textwidth]{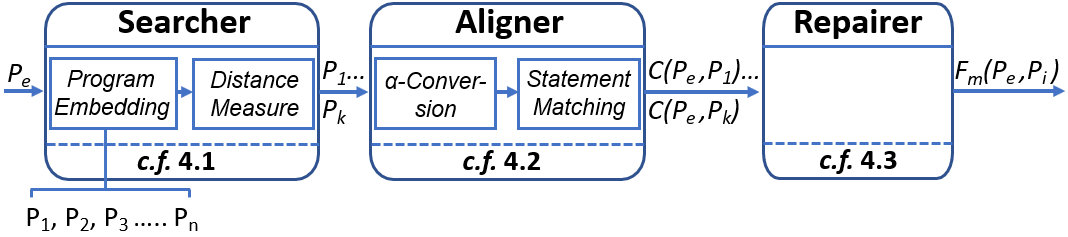}
	\caption{The overall workflow of the $\tool$ system.}
	\label{fig:arc}
\end{figure}

We now briefly present an overview of the workflow of the $\tool$
system. The three major components are outlined in
Figure~\ref{fig:arc}: (1) \textit{Searcher}, (2) \textit{Aligner}, and
(3) \textit{Repairer}.

\begin{enumerate}[label*=(\arabic*)] 
%
		
	\item \textbf{Searcher}: The Searcher component takes as input
          an incorrect program $P_{e}$ and searches for the top $k$
          closest solutions among all the correct programs $P_{1},
          P_{2}, P_{3},...$ and $P_{n}$ for the given exercise. The
          key challenge is to search a large number of programs in an
          efficient and precise manner. The Searcher component
          consists of:
	\begin{itemize} 
		\item \textbf{Program Embedder}: The Program Embedder
                  converts $P_{e}$ and $P_{1}, P_{2}, P_{3},...P_{n}$
                  into numerical vectors. We propose a new scheme of
                  embedding programs that improves the original
                  characteristic vector representation used previously
                  in~\citet{jiang2007deckard}.
		
		\item \textbf{Distance Calculator}: Using program
                  embeddings, the Distance Calculator computes the top
                  $k$ closest reference solutions in the Euclidean
                  space. The advantage is such distance computations are much more
                  scalable than the tree edit distance algorithm.
	\end{itemize}

	\item \textbf{Aligner}: After computing a set of top $k$
          candidate programs for comparison, the Aligner component
          aligns each of the candidate programs $P_{1},...P_{k}$ \wrt
          $P_{e}$. 
          
          	\begin{itemize} 
		\item \textbf{$\alpha$-conversion}: We propose a 
		usage-set based $\alpha$-con\-version. Particularly we 
		profile each variable on its usage and summarize it into  
		one single numeric vector via our new embedding scheme. 
		Next we minimize the distance between two sets of variables 
		based on the Euclidean distance 
		metric. This novel technique not only enables efficient 
		computation but also achieves high accuracy. 
		\item \textbf{Statement Matching}: After the $\alpha$-conversion, 
		we align basic blocks, and in turn individual statements within 
		each aligned basic blocks.
	\end{itemize}
	
	 Based on the alignment, we produce a set of
          syntactical discrepancies $\mathcal{C}(P_{e},P_{k})$. This
          is an important step as misaligned programs
          would result in an imprecise set of corrections.
	
%
%
	
	\item \textbf{Repairer}: Given $\mathcal{C}(P_{e},P_{k})$,  the 
	Repairer component produces a set of fixes $\mathcal{F}(P_{e},P_{k})$. 
	Later it minimizes $\mathcal{F}(P_{e},P_{k})$ by removing the
          syntactic/semantic redundancies that are unrelated to the
          root cause of the error. We propose our minimization
          technique based on an optimized dynamic analysis to make the minimal repair computation
          efficient and scalable.
		
\end{enumerate}

\section{The Search, Align, and Repair Algorithm}
\label{section:method}

This section presents our feedback generation algorithm. 
In particular, it describes the three key functional components: 
\textit{Search}, \textit{Align}, and \textit{Repair} to cope 
with the challenges discussed in Section~\ref{section:intr}.

\subsection{Search}	
To realize our goal of using correct programs to repair an incorrect 
program, the very first problem we need to solve is to identify 
a small subset of correct programs among thousands of submissions that are 
relevant to fixing the incorrect program in an efficient and precise manner. 
To start with, we perform exact matching between reference solutions 
and the incorrect program \wrt their control-flow structures.  

\begin{definition}{(Control-Flow Structure)}	
        Given a program $P$, its control-flow
        structure, $\mathit{CF}(P)$, is a syntactic 
	registration of how control statements in $P$ are 
	coordinated. For brevity, we 
	denote $\mathit{CF}(P)$ to simply be a sequence of 
	symbols (\eg $\mathit{For}_{\mathit{start}}$, 
	$\mathit{For}_{\mathit{start}}$, 
	$\mathit{If}_{\mathit{start}}$, 
	$\mathit{If}_{\mathit{end}}$, 
	$\mathit{Else}_{\mathit{start}}$, 
	$\mathit{Else}_{\mathit{end}}$, 
	$\mathit{For}_{\mathit{end}}$, 
	$\mathit{For}_{\mathit{end}}$ for the program 
	in Figure~\ref{fig:ic1}).
\end{definition}

Given the selected programs with the same control-flow
structure, we search for similar programs 
using a syntactic approach (in contrast to a dynamic approach) for two reasons:
(1) less overhead (\ie faster) and (2) more fault-tolerant 
as runtime dynamic 
\textit{traces} likely lead to greater 
differences if students made an error especially on 
control predicates. However, using the standard tree edit 
distance~\cite{tai1979tree} as the distance measure does 
not scale to a large number of programs. We propose a new 
method of embedding ASTs into numerical vectors,
namely the position-aware characteristic vectors, with 
which Euclidean distance can be computed to represent 
the syntactic distance. Next, we briefly 
revisit Definitions~\ref{def:pat} and~\ref{def:cv} 
proposed in~\cite{jiang2007deckard}, upon which our new 
embedding is built.

Given a binary tree, we define a family of atomic 
tree patterns to capture structural information of 
a tree. They are parametrized by a parameter q, the 
height of the patterns. 

\begin{definition}{($q$-Level Atomic Tree Patterns)}
\label{def:pat}	
A $q$-level atomic pattern is a 
complete binary tree of height $q$. Given a 
label set $L$, including the empty label 
$\epsilon$, there are at most $\vert L \vert^{2^q-1}$ 
distinct $q$-level atomic patterns.
\end{definition}

\begin{definition}{($q$-Level Characteristic Vector)}	
\label{def:cv}	
	Given a tree $T$, its $q$-level 
	characteristic vector is 
	$\langle b_{1},  ...., b_{\vert L \vert^{2^q-1}} \rangle$, where $b_{i}$ 
	is the number of occurrences of the $i$-th $q$-level 
	atomic pattern in $T$.
\end{definition}

\begin{figure}[t!]
	\begin{center}
		\includegraphics[width=0.46\textwidth]{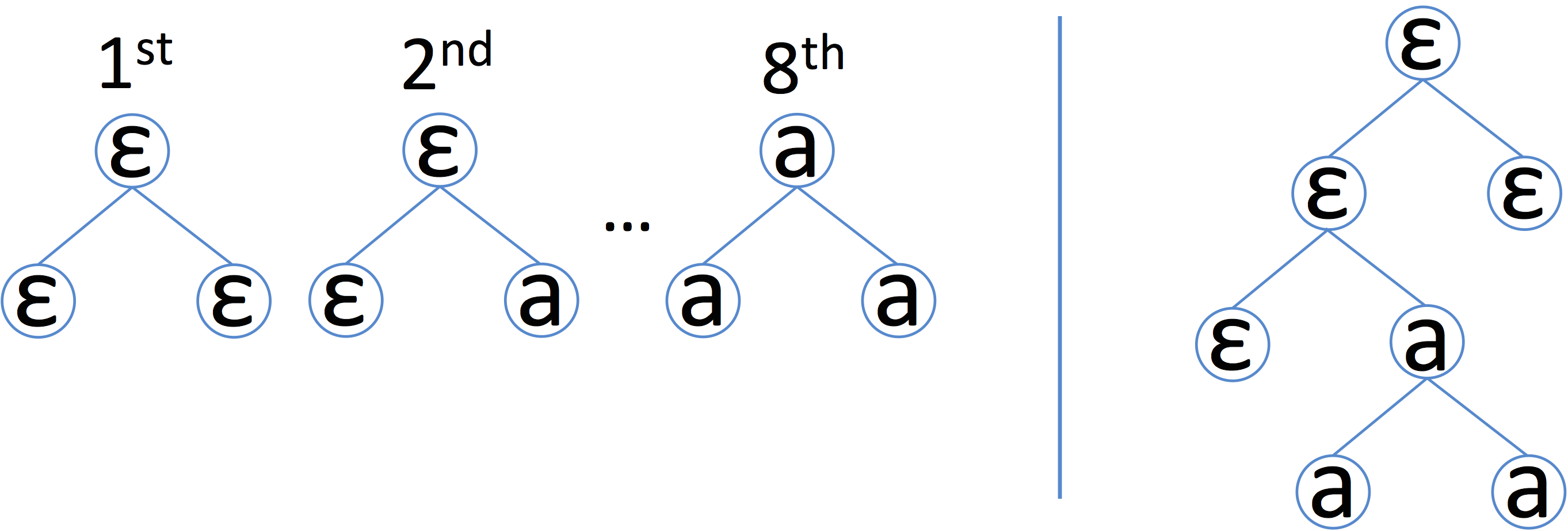}
		\caption{The eight 2-level atomic tree patterns for a 
			label set of $\{a,\epsilon\}$ on the left. 
			Using 2-level characteristic vector to embed 
			the tree on the right: 
			$\langle1,1,0,0,0,0,0,1\rangle$; 
			and 2-level \pavector: 
			$\langle \langle4\rangle, \langle3\rangle,....,\langle2\rangle \rangle$.}
		\label{fig:pe}
	\end{center}
\end{figure}

Figure~\ref{fig:pe} depicts 
an example. Given the label set $\{a,\epsilon\}$, 
the tree on the right can be embedded into 
$\langle1,1,0,0,0,0,0,1\rangle$ using 2-level 
characteristic vectors. The benefit of such embedding 
schemes is the realization of gauging program 
similarity using \textit{Hamming distance}/\textit{Euclidean distance} 
metric which is much faster to compute. In order to further enhance 
the precision of the AST embeddings, we introduce a 
\emph{position-aware characteristic vector} embedding which incorporates 
more structural information into the encoding vectors.

\begin{definition}{($q$-Level Position-Aware Characteristic Vector)}	
	\label{def:pacv}	
	Given a tree $T$, its $q$-level 
	position-aware characteristic vector is 
	$\langle \langle b_{1}^{1},....,b_{1}^{n_{1}}\rangle, 
	\langle b_{2}^{1},....,b_{2}^{n_{2}} \rangle, 
	\langle b_{\vert L \vert^{2^q-1} }^{L},....,
	b_{\vert L \vert^{2^q-1} }^{n_{\vert L\vert^{2^q-1}}}\rangle  
	\rangle$, \\ where $b_{i}^{j}$ is the height of the root node of 
	the $j$-th $i$-th $q$-level atomic pattern in $T$.	
\end{definition} 

Essentially we expand each element $b_{i}$ defined in the $q$-level 
characteristic vector into a list of heights for each 
$i$-th $q$-level atomic pattern in 
$T$ ($b_{i} = \vert \langle b_{i}^{1},....,b_{i}^{n_{i}}   
\rangle \vert$). 
Using a 2-level \pavector, the same tree in Figure~\ref{fig:pe} 
can be embedded into $\langle \langle4\rangle, \langle3\rangle
,....,\langle2\rangle \rangle$. For brevity, we shorten the $q$-level 
position-aware characteristic vector to be 
$\langle h_{b_{1}},....,h_{b_{\vert L \vert^{2^q-1}}} \rangle$ 
where $h_{b_{i}}$ represents the vector of heights of all $i$-th 
$q$-level atomic tree patterns. The distance
between two $q$-level \pavector is:

\begin{align}
& \sqrt{\sum\limits_{i=1}^{\vert L \vert^{2^q-1}} {\vert\vert \mathit{sort}(h_{b_{i}}, \phi) - \mathit{sort}(h_{b_{i}}^\prime, \phi) \vert\vert_{2}^{2} }} \\ 
& \mbox{where} \; \phi = \mathit{max}(\vert h_{b_{i}}\vert, \vert h_{b_{i}}^\prime \vert)
\end{align}

where $\mathit{sort}(h_{b_{i}}, \phi)$ means sorting 
$h_{b_{i}}$ in descending order followed 
by padding zeros to the end if $h_{b_{i}}$ 
happens to be the smaller vector of the two. 
The idea is to normalize $h_{b_{i}}$ 
and $h_{b_{i}}^\prime$ prior to the distance
calculation. $\vert\vert ... \vert\vert_{2}^{2}$ 
denotes the square of the L2 norm. We are given the incorrect program 
(having $m$ nodes in its AST) and $\uprho$ 
candidate solutions (assuming each has the same 
number of nodes $n$ in their ASTs to simplify the 
calculation). Creating the embedding as well as 
computing the Euclidean distance on the resulting 
vectors has a worst-case complexity of 
$O(m+\uprho*n+\uprho*(\vert h_{b_{1}} \vert,...., \vert h_{b_{\vert L \vert^{2^q-1}}}\vert))$. 
Because the {\pavector}s can be computed offline 
for the correct programs, we can further reduce 
the time complexity to 
$O(m+\uprho*(\vert h_{b_{1}} \vert,...., \vert h_{b_{\vert L \vert^{2^q-1}}}\vert))$. 
In comparison, the state-of-the art 
Zhang-Shasha tree edit distance algorithm~
\cite{zhang1989simple} runs in $O(\uprho*m^{2}n^{2})$. 
Our large-scale evaluation shows that this new 
program embeddings using position-aware characteristic 
vectors and the new distance measure not only leads to significantly 
faster search than tree edit distance on ASTs but also 
negligible precision loss. 

\subsection{Align}
\label{subsection:align}
The goal of the align step is to compute 
Discrepancies (Definition~\ref{def:dis}). 
The rationale is that after a syntactically similar 
program is identified, it must be correctly 
aligned with the incorrect program such that the 
differences between the aligned statements 
can suggest potential corrections.

\begin{definition}{(Discrepancies)}	
	\label{def:dis}	
Given the incorrect program $P_{e}$ and a correct 
program $P_{c}$, discrepancies, denoted by $\mathcal{C}({P}_{e},{P}_{c})$,
is a list of pairs, $({S}_{e}$,${S}_{c})$, where 
${S}_{e}$/${S}_{c}$ is a non-control statement\footnote{Hereinafter we denote non-control statement to include loop headers, branch conditions, \etc} 
in ${P}_{e}$/${P}_{c}$. 
\end{definition} 

Aligning a reference solution with the 
incorrect program is a crucial step in 
generating accurate fixes, and in turn feedback. 
Figure~\ref{fig:align} shows an example, in which 
the challenges that need to be addressed are: 
(1) renaming $blackWhite$ in the correct solution 
to $ch$ --- failing to do so will result in incorrect 
let alone precise fix; (2) computing the correct alignment 
which leads to the minimum fix of inserting $ch = !ch$ on 
line 12 in Figure~\ref{fig:ainc2}; and (3) 
solving the previous two tasks in a timely manner to 
ensure a good user experience. Our key idea of solving 
these challenges is to reduce the alignment problem to 
a distance computation problem, specifically, of finding 
an alignment of two programs that minimizes 
their syntactic distances. We realize this 
idea in two-steps: variable-usage based $\alpha$-conversion 
and two-level statement matching.

\begin{figure}[h]
	\begin{subfigure}[b]{0.23\textwidth}
		\lstset{style=mystyle}
\begin{lstlisting}[basicstyle=\fontsize{5}{8.8}\ttfamily]
static void Main(string[] args){
    int length = 8;
    @bool ch = true;@

    for (int i = 0; i < length; i++){ 
        for (int j = 0; j < length; j++){ 
            if (@ch@) Console.Write("X");
            else Console.Write("O");

            @ch = !ch;@
        }
        /* inserting ch =!ch here*/
        Console.Write("\n");
    }}
\end{lstlisting}		
		\caption{Incorrect program in Fig~\ref{fig:ic2}.}
		\label{fig:ainc2}
	\end{subfigure}
	\;  
	\begin{subfigure}[b]{0.23\textwidth}
		\lstset{style=mystyle}
\begin{lstlisting}[basicstyle=\fontsize{5}{8.8}\ttfamily]
static void Main(string[] args){
    @bool blackWhite = true;@
    for (int i = 0; i < 8; i++){
        for (int j = 0; j < 8; j++){        
            if (@blackWhite@)
                Console.Write("X");
            else
                Console.Write("O");
            @blackWhite = !blackWhite;@
        }

        @blackWhite = !blackWhite;@
        Console.WriteLine("");
    }}
\end{lstlisting}		    
		\caption{Correct program in Fig~\ref{fig:c2}.}
		\label{fig:acor2}
	\end{subfigure}
	\caption{Highlighting the usage of variable $ch$ and $blackWhite$ for aligning programs in Figures~\ref{fig:ic2} and~\ref{fig:c2}.}
	\label{fig:align}
\end{figure}


\paragraph{\textbf{Variable-usage based $\alpha$-Conversion}}
The dynamic app\-roach of comparing the runtime traces for each 
variable suffers from the same scalability issues 
mentioned in the search procedure. Instead, we present a syntactic approach --- \textit{variable-usage} 
based $\alpha$-conversion. Our intuition is 
that how a variable is being used in a program serves as 
a good indicator of its identity. To this end, we represent 
each variable by profiling its usage in the program. 

\begin{definition}{(Usage Set)}	
	\label{def:us}	
	Given a program $P$ and the variable set 
	$\mathit{Vars}(P)$, a usage set of a variable 
	$v \in \mathit{Vars}(P)$ consists of all the 
	non-control statements featuring $v$ in $P$.	
\end{definition} 

We then collect the usage set for each 
variable in $P_{e}/P_{c}$ to form 
$\mathcal{U}({P}_{e})/\mathcal{U}({P}_{c})$. 
Now the goal is to find a one-to-one 
mapping between $\mathit{Vars}(P_{e})$ and 
$\mathit{Vars}(P_{c})$ which minimizes the 
distance between $\mathcal{U}({P}_{e})$ and 
$\mathcal{U}({P}_{c})$.

\begin{equation}
\label{equ:argmin}
	\mbox{$\alpha$-\textit{conversion}}\; = \argminA_{ \Scale[0.5]{Vars(P_{c}) \leftrightarrow  Vars(P_{e})}} \Delta(\mathcal{U}({P}_{c}), \mathcal{U}({P}_{e})) 
\end{equation}

We can now compute the tree-edit distance between 
statements in any two usage sets in $\mathcal{U}({P}_{e})$ 
and $\mathcal{U}({P}_{c})$, and then find the 
matching that adds up to the smallest distance 
between $\mathcal{U}({P}_{e})$ 
and $\mathcal{U}({P}_{c})$. However, the total 
number of usage sets in $\mathcal{U}({P}_{e})$ 
and $\mathcal{U}({P}_{c})$ (denoted by the level of 
usage set) and the number of statements in each usage 
set (denoted by the level of statement) will lead 
this approach to a combinatorial explosion 
that does not scale in practice. Instead, we rely 
on the new program embeddings to represent each 
usage set with only one single position-aware characteristic 
vector. Using $H_{v_{i}}$/$H_{v_{i}}^\prime$ to 
denote the vector for usage set of $v_{i} \! \in \! \mathit{Vars}(P_{e})$/
$v_{i}^\prime \!\in\! \mathit{Vars}(P_{c})$, we instantiate 
Equation~\ref{equ:argmin} into

\begin{align}
& \mbox{$\alpha$-\textit{conversion}}\; = \argminA_{ \scriptstyle{ v_{i} \leftrightarrow v_{i}^\prime} } \sum_{i=1}^{\omega} \vert\vert H_{v_{i}} - H_{v_{i}}^\prime \vert\vert_{2} \label{equ:argminv} \\ 
& \mbox{where} \; \omega = \mathit{min}(\vert \mathit{Vars}(P_{c}) \vert, \vert \mathit{Vars}(P_{e}) \vert)
\end{align}

The benefits of this instantiation are: (1) it only 
focuses on the combination at the level of usage set, and 
therefore eliminates a vast majority of 
combinations at the level of statement; and (2) it can quickly compute the Euclidean distance between two 
usage sets. Note the computed mapping between 
$\mathit{Vars}(P_{c})$ and $\mathit{Vars}(P_{e})$ in Equation~\ref{equ:argminv} 
does not necessarily lead to the correct $\alpha$-conversion. 
To deal with the phenomenon that variables may 
need to be left unmatched if their usages are too 
dissimilar, we apply the Tukey's method
~\cite{tukey1977exploratory} to remove statistical 
outliers based on the Euclidean distance. After the 
$\alpha$-conversion, we turn ${P}_{c}$ into 
${P}_{\alpha\!c}$. 

\begin{figure}[t!]
	\centering
	\includegraphics[width=0.47\textwidth]{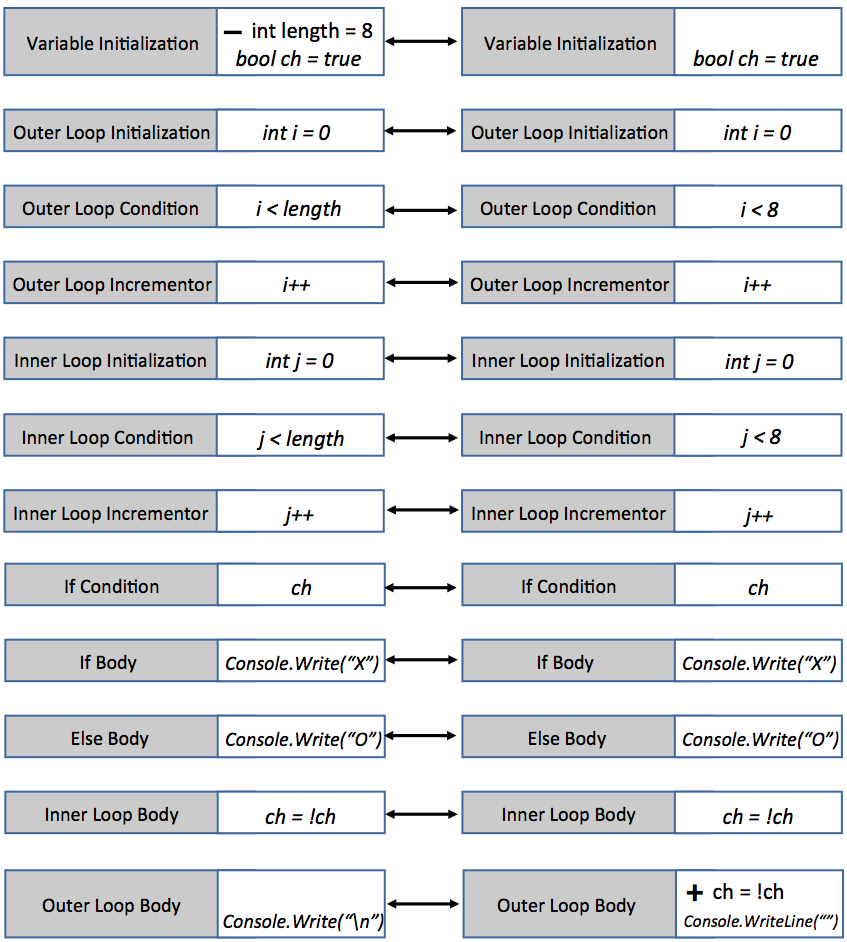}
	\caption{Aligning basic blocks for programs in Figure~\ref{fig:ic2} and~\ref{fig:c2} after 
		$\alpha$-conversion (aligned blocks are connected by arrows); matching statements within each pair of aligned basic blocks (
		italicized statements denotes matching; statements annotated by +/- denotes insertion/deletion.).}
	\label{fig:alignres}
\end{figure}

\paragraph{\textbf{Two-Level Statement Matching}}
We leverage program 
str\-ucture to perform statement matching at two 
levels. At the first level, we fragment $P_{e}$ and 
$P_{\alpha\!c}$ into a 
collection of basic blocks according to their control-flow
structure, and then align each pair in order. This alignment 
will result in a 1-1 mapping of the basic blocks 
due to $\mathit{CF}(P_{e}) \equiv \mathit{CF}({P}_{\alpha\!c})$. 
Next, we match statements/expressions 
only within aligned basic blocks. 
In particular, we pick the matching that 
minimizes the total syntactic distance (tree edit distance) 
among all pairs of matched statements within each aligned 
basic blocks. Figure~\ref{fig:alignres} depicts the result 
of aligning programs in Figure~\ref{fig:align}.

\begin{figure*}[htb!]
	\begin{subfigure}[b]{0.48\textwidth}
		\begin{subfigure}[b]{0.48\textwidth}
			\lstset{style=mystyle}
\begin{lstlisting}[numbers=none,basicstyle=\fontsize{5}{8.8}\ttfamily]
static void Main(string[] args){
    for(int i = 0;i < 8;i++){
        String chess = @""@;
        for (int j = 0;j < 8;j++){
            if (j % 2 == 0)
                chess += "X";
            else
                chess += "0";
        }
        @Console.Write(chess+"\n");@
    }}

\end{lstlisting}		
		\end{subfigure}
		\;  
		\begin{subfigure}[b]{0.48\textwidth}
			\lstset{style=mystyle}
\begin{lstlisting}[numbers=none,basicstyle=\fontsize{5}{8.8}\ttfamily]
static void Main(string[] args){
    for (int i = 0; i < 8; i++){
       String query = @string.Empty@;
       for (int j = 0; j < 8; j++){
            if ((i + j) % 2 == 0)
                query += "X";
            else
                query += "O";
       }
       @Console.WriteLine(query);@
    }}

\end{lstlisting}		    
		\end{subfigure}		
	\caption{Syntactic differences between programs in Figures~\ref{fig:ic1} and~\ref{fig:c1}.}
	\label{fig:diff1}	
	\end{subfigure}
	\;
	\begin{subfigure}[b]{0.48\textwidth}
		\begin{subfigure}[b]{0.48\textwidth}
			\lstset{style=mystyle}
\begin{lstlisting}[numbers=none,basicstyle=\linespread{.65}\fontsize{5}{8.8}\ttfamily]
static void Main(string[] args){
    for (int i = @1@; i < @9@; i++){
        for (int j = @1@; j < @9@; j++){
            if (i % 2 > 0)
                if (j % 2 @>@ 0)
                    Console.Write(@"O"@);
                else
                    Console.Write(@"X"@);
            else
                if (j % 2 @>@ 0)
                    Console.Write(@"X"@);
                else
                    Console.Write(@"O"@);
        }
        Console.WriteLine();
    }}

\end{lstlisting}		
		\end{subfigure}
		\;  
		\begin{subfigure}[b]{0.48\textwidth}
			\lstset{style=mystyle}
\begin{lstlisting}[numbers=none,basicstyle=\linespread{.65}\fontsize{5}{8.8}\ttfamily]
static void Main(string[] args){
    for (int i = @0@; i < @8@; i++){
        for (int j = @0@; j < @8@; j++){
            if (i % 2 == 0)
                if (j % 2 @==@ 0)
                    Console.Write(@"X"@);
                else
                    Console.Write(@"O"@);
            else
                if (j % 2 @==@ 0)
                    Console.Write(@"O"@);
                else
                    Console.Write(@"X"@);
        }
        Console.WriteLine();
    }}

\end{lstlisting}		    
		\end{subfigure}		
	\caption{Semantic differences between programs in Figures~\ref{fig:ic3} and~\ref{fig:c3}.}
	\label{fig:diff2}	
	\end{subfigure}

	\caption{Syntactic and semantic differences.}
\end{figure*}

\subsection{Repair}
\label{subsection:repair}
Given $\mathcal{C}({P}_{e},{P}_{\alpha\!c})$, 
we generate 
$\mathcal{F}({P}_{e}$, ${P}_{\alpha\!c})$ 
as follows:

\begin{itemize}
	\item \textbf{Insertion Fix}: Given a pair $({S}_{e}$,${S}_{c})$, 
	if ${S}_{e} = \emptyset$ meaning ${S}_{c}$ is not aligned to any 
	statement in ${P}_{e}$, an insertion operation will be produced 
	to add ${S}_{c}$.
	
	\item \textbf{Deletion Fix}: Given a pair $({S}_{e}$,${S}_{c})$, 
	if ${S}_{c} = \emptyset$ meaning ${S}_{e}$ is not aligned to any 
	statement in ${P}_{c}$, an deletion operation will be produced 
	to delete ${S}_{e}$.
	
	\item \textbf{Modification Fix}: Given a pair $({S}_{e}$,${S}_{c})$, 
	if ${S}_{c} \ne \emptyset$ and ${S}_{e} \ne \emptyset$, a modification 
	operation will be produced consisting the set of standard tree 
	operations to turn the AST of ${S}_{e}$ to that of ${S}_{c}$.
\end{itemize}

The computed set of potential fixes $\mathcal{F}({P}_{e}$, ${P}_{\alpha\!c})$ typically contains a large set of redundancies 
that are unrelated to the root cause of the 
error in $P_{e}$. Such redundancies can be classified 
into two categories:
\begin{itemize}
	\item \textbf{Syntactic Differences:} 
	${P}_{e}$ and ${P}_{\alpha\!c}$ 
	may use different expressions in 
	method calls, parameters, constants, \etc 
	Figure~\ref{fig:diff1} highlights such syntactic differences. 	
	\item \textbf{Semantic Differences:} 
	More importantly, 
	$\mathcal{F}({P}_{e}$, ${P}_{\alpha\!c})$ 
	may also contain semantically-equivalent differences such as different conditional statements 
	yet logically 
	equivalent $\mathit{if}$ statement; different initialization 
	and conditional expressions that evaluate to the 
	same number of loop iterations; or expressions 
	computing similar values composed of different variables 
	(not variable names). Figure~\ref{fig:diff2} highlights such semantic differences.
\end{itemize}

The Repair component is responsible for 
filtering out all those benign differences 
that are mixed in 
$\mathcal{F}({P}_{e}$, ${P}_{\alpha\!c})$, 
in other words, picking 
$\mathcal{F}_{m}({P}_{e}, {P}_{\alpha\!c})$ 
($\mathcal{F}_{m}({P}_{e}, {P}_{\alpha\!c})  \subseteq  \mathcal{F}({P}_{e}, {P}_{\alpha\!c})$)
that only fix the 
issues in $P_{e}$ while leaving the correct parts unchanged. 
A direct brute-force approach is to exhaustively incorporate each subset of 
$\mathcal{F}({P}_{e}$, ${P}_{\alpha\!c})$ into the original 
incorrect program, and test the resultant program 
by dynamic execution. This approach yields 
$2^{\vert \mathcal{F}({P}_{e}, {P}_{\alpha\!c}) \vert} - 1$ 
number of trials in total. As the number of operations in 
$\mathcal{F}({P}_{e}$, ${P}_{\alpha\!c})$ 
increases, the search space become intractable. Instead, 
we apply several optimization techniques to make 
the procedure more efficient and scalable (\cf Section~\ref{section:eva}).

\subsection{The Repair Algorithm in $\tool$}

We now present the complete repair algorithm in $\tool$. The system 
first matches the control flow (Line~\ref{line:cls}-\ref{line:cle}); 
then selects top $k$ candidates for repair (Line~\ref{line:diss}-\ref{line:dise}); 
Line~\ref{line:sas}-\ref{line:da} denotes the search, align and repair 
procedure, whereas Line~\ref{line:ft} generates the feedback.


\newcommand\mycommfont[1]{\footnotesize\ttfamily{#1}}
\SetCommentSty{mycommfont}


\begin{algorithm}[h]
	\DontPrintSemicolon 
	\SetKwProg{proc}{function}{}{}
	
	\SetKwFunction{procname}{FeedbackGeneration}
	
	\SetKwData{FB}{$f$}
	\SetKwData{FL}{$\ell$}
	\SetKwData{Incor}{${P}_{e}$}
	\SetKwData{Refer}{${P}_{s}$}
	\SetKwData{CL}{${P}_{cs}$}
	\SetKwData{CAs}{${P}_{dis}$}
	\SetKwData{CA}{${P}_{c}$}
	\SetKwData{alCA}{${P}_{\alpha\!c}$}
	\SetKwData{Dis}{$\mathcal{C}({P}_{e}$, ${P}_{\alpha\!c})$}	
	\SetKwData{Fix}{$\mathcal{F}({P}_{e}$, ${P}_{\alpha\!c})$}
	\SetKwData{Min}{$\mathcal{F}_{m}({P}_{e}$, ${P}_{\alpha\!c})$}
	
	\tcc{\Incor: an incorrect program; \Refer: all correct solutions; \FL: feedback level}			
	\proc{\procname{\Incor, \Refer, \FL}}{ 
		\Begin{
			
			\tcc{identify ${P}_{cs}$ that have the same control flow of ${P}_{e}$ among all solutions ${P}_{s}$}			
			\CL $\gets$ $\emptyset$ \label{line:cls} \\
			\For{$P$ $\in$ \Refer}{
				\If{CF(${P}$) = CF(\Incor)}{
					\CL $\gets$ \{ ${P}$ \} $\cup$ \CL \label{line:cle} 
				}			
			}
			
			\tcc{collect ${P}_{dis}$ that consist of k most similar programs with ${P}_{e}$.}						
			\CAs $\gets$ $\emptyset$ \label{line:diss} \\
			\For{$P$ $\in$ \CL}{
			   \If{$\mathcal{D}$($P$, \Incor) < $\mathcal{D}$($P_{kth} \in \CAs$, \Incor)}
			   {
			   	    \CAs $\gets$ \CAs $\setminus$ \{ $P_{kth}$ \}  	\\		   	    
			   	    \CAs $\gets$ \{ $P$ \} $\cup$ \CAs  \label{line:dise}
			   }
			}
					
			$n  \gets \infty$ \\ 		
			$\mathcal{F}_{m}({P}_{e}) \gets$ null \tcp*{minimum set of fix for $P_{e}$}
				
			\For{\CA $\in$ \CAs}{
				
				\alCA $\gets$ $\alpha$-Conversion(\Incor, \CA) \label{line:sas} \\
				\Dis $\gets$ Discrepencies(\Incor, \alCA) \label{line:sae}\\
				\Fix $\gets$ Fixes(\Dis) \label{line:sae}\\				
				\Min $\gets$ Minization(\Fix)  \label{line:da}
				
				\If{$\vert \Min \vert$ < $n$}{
					$\mathcal{F}_{m}({P}_{e}) \gets$ \Min  \\
					$n \gets \vert \Min \vert$
				}
				
			}
			
			\tcc{translate fixes into feedback message.}						
			\FB = Translating($\mathcal{F}_{m}({P}_{e})$, \FL)  \label{line:ft}  \\
			\Return{\FB}		
		}
	}
	
	\caption{$\tool$'s feedback generation.}
	\label{alg:fg}
\end{algorithm}


\section{Implementation and Evaluation}
\label{section:eva}

We briefly describe some of the implementation details of $\tool$, and
then report the experimental results on benchmark problems.  We also
conduct an in-depth analysis for measuring the usefulness of the
different techniques and framework parameters, and conclude with an
empirical comparison with the CLARA tool~\cite{gulwani2016automated}.

\subsection{Implementation}
$\tool$ is implemented in \Csharp. We use 
the Microsoft Roslyn compiler framework for parsing ASTs and dynamic execution. 
We keep 5 syntactically most similar programs 
as the reference solutions. For the printing problem 
from Microsoft-DEV204.1X, we convert the console operation to 
string operation using the $\mathit{StringBuilder}$ 
class. As for all the program embeddings, we 
use 1-level characteristic vectors. All experiments 
are conducted on a Dell XPS 8500 with a 3rd Generation 
Intel Core\textregistered$^{\text{TM}}$ i7-3770 
processor and 16GB RAM.

\begin{table*}[htbp!]
	\begin{center}
		\begin{adjustbox}{max width=.8\textwidth}
			\begin{tabular}{|c | c | c | c | c | c | c | c | c} 
				\hline
				\textbf{Benchmark} & \begin{tabular}{@{}c@{}}\textbf{Average} \\ \textbf{(LOC)}\end{tabular} & \begin{tabular}{@{}c@{}}\textbf{Median} \\ \textbf{(LOC)}\end{tabular} & 
				\begin{tabular}{@{}c@{}}\textbf{Correct} \\ \textbf{Attempts}\end{tabular}  & \begin{tabular}{@{}c@{}}\textbf{InCorrect} \\ \textbf{Attempts}\end{tabular} & 
				\begin{tabular}{@{}c@{}}\textbf{Generated Feedback} \\ \textbf{on Minimum Fixes}\end{tabular} & 
				\begin{tabular}{@{}c@{}}\textbf{Average Time} \\ \textbf{Time (in s)}\end{tabular} & 
				\begin{tabular}{@{}c@{}}\textbf{Median Time} \\ \textbf{Time (in s)}\end{tabular} \\ 
				\hline
				\hline
				Divisibility & 4 & 3 & 609 &398  &379 (\%95.2) & 0.84 & 0.89 \\						
				\hline
				ArrayIndexing & 3 & 3 & 524 & 449  & 421 (\%93.8) & 0.82 & 0.79 \\						
				\hline
				StringCount & 6 & 4 & 803 & 603  & 562 (\%93.2) & 1.38 & 1.13 \\						
				\hline
				Average & 8 & 7 & 551 & 465  & 419 (\%90.1) & 1.16 & 1.08 \\						
				\hline
				ParenthesisDepth & 18 & 15 & 505 & 315  & 277 (\%88.0) & 2.25 & 2.32 \\								
				\hline
				Reversal & 14 & 11 & 623 & 398  & 374 (\%94.0) & 2.10 & 1.77 \\						
				\hline
				LCM & 15 & 10 & 692 & 114  & 97(\%85.1) & 2.07 & 2.14 \\						
				\hline
				MaximumDifference & 11 & 8 & 928 & 172  & 155 (\%90.1) & 1.40 & 1.78 \\						
				\hline
				BinaryDigits & 10 & 7 & 518 & 371  & 343 (\%92.4) & 2.29 & 2.01 \\						
				\hline
				Filter & 12 & 8 & 582 & 88 & 71 (\%80.7) & 1.64 & 1.72 \\						
				\hline
				FibonacciSum & 14 & 12 &899 &131 & 114 (\%87.0) & 2.78 & 2.45 \\						
				\hline
				K-thLargest & 11 & 9 & 708 & 241  & 208 (\%86.3) & 1.51 & 1.16 \\						
				\hline
				SetDifference & 16 & 13 & 789 & 284  & 236 (\%83.1) & 1.68 & 1.19 \\						
				\hline
				Combinations & 14 & 10 & 638 & 68  &56 (\%82.4) & 2.13 & 1.94 \\						
				\hline
				MostOnes & 29 & 33 & 748 &394   &328 (\%87.9) & 2.72 & 2.15 \\						
				\hline
				ArrayMapping & 16 & 14 & 671 & 315  & 271 (\%86.0) & 1.46 & 1.8 \\						
				\hline
				\hline
				Printing & 24 & 21 &2281  &742   &526 (\%70.9)  & 3.16 & 2.77 \\						
				\hline
			\end{tabular}
		\end{adjustbox}
	\end{center}
	\caption{Experimental results on the benchmark problems obtained from CodeHunt and edX course.}
	\label{Table:res}
\end{table*}

\subsection{Results}
We evaluate $\tool$ on the chessboard printing problem 
from Microsoft-DEV204.1X as well as 16 out of the 24 problems 
(the other seven are rarely attempted by students) on the 
CodeHunt education platform (ignoring the submissions that 
are syntactically incorrect).\footnote{Please refer to the 
supplementary material for a brief description of each 
benchmark problem.}
Table~\ref{Table:res} shows the results. Overall, $\tool$ 
generates feedback based on minimum fixes for 4,311 submissions 
out of 4,806 incorrect programs in total ($\apeq$ 90\%) within 
two seconds on average. Our evaluation results validate the 
assumption we made since all the minimal fixes modify
up to three lines only. Another interesting finding is that $\tool$ performed better 
on CodeHunt programming submissions than on \course's 
assignment despite the fewer number of correct programs. 
After further investigation, we conclude that the most likely 
cause for this is the printing nature of \course's exercise 
placing little constraint on the control-flow structure. In 
an extreme, we find students writing straight-line programs 
even in several different ways, and therefore those programs 
are more diverse and difficult for $\tool$ to precisely find the 
closest solutions. On the other hand, CodeHunt programs are 
functional and more structured. Even though there are fewer reference 
implementations, $\tool$ is capable of repairing 
more incorrect submissions.

%
%
%
%

 \begin{figure*}[!ht]
	\begin{subfigure}[b]{0.38\textwidth}
		\includegraphics[width=\textwidth]{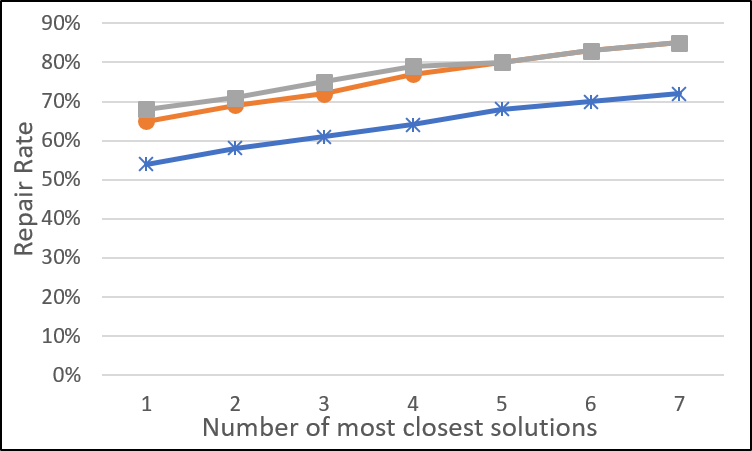}
		\caption{Comparison on capability.}
	\end{subfigure}
	\begin{subfigure}[b]{0.19\textwidth}
		\includegraphics[width=\textwidth]{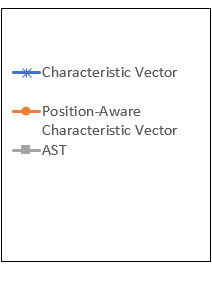}
	\end{subfigure}	
	\begin{subfigure}[b]{0.38\textwidth}
		\includegraphics[width=\textwidth]{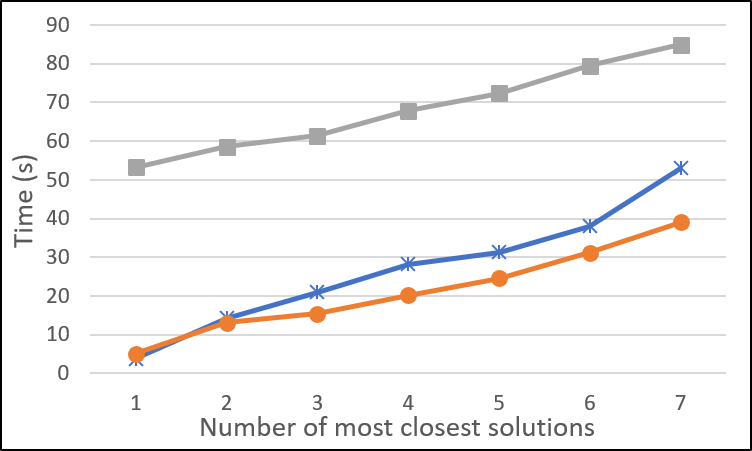}
		\caption{Comparison on performance.}
	\end{subfigure}
	\caption{Position-Aware Characteristic Vector vs. Characteristic Vector.}
	\label{fig:vectors}
\end{figure*}

 \begin{figure*}[!ht]
	\begin{subfigure}[b]{0.38\textwidth}
		\includegraphics[width=\textwidth]{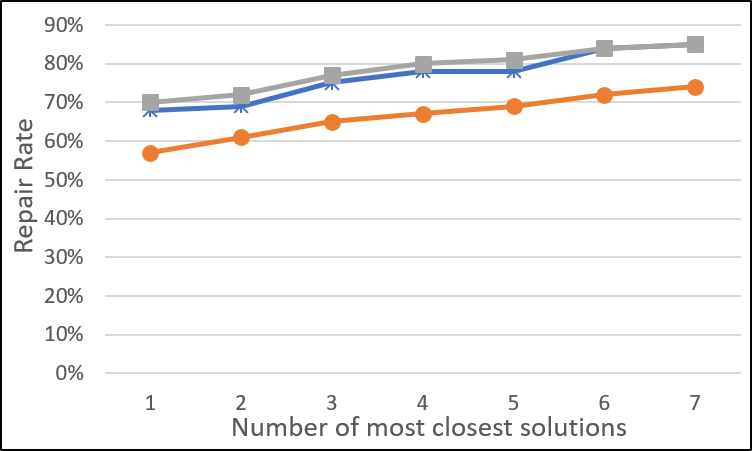}
		\caption{Comparison on capability.}
	\end{subfigure}
	\begin{subfigure}[b]{0.19\textwidth}
		\includegraphics[width=\textwidth]{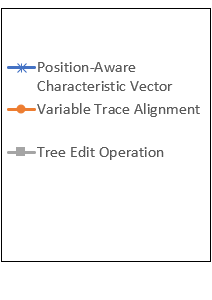}
	\end{subfigure}	
	\begin{subfigure}[b]{0.38\textwidth}
		\includegraphics[width=\textwidth]{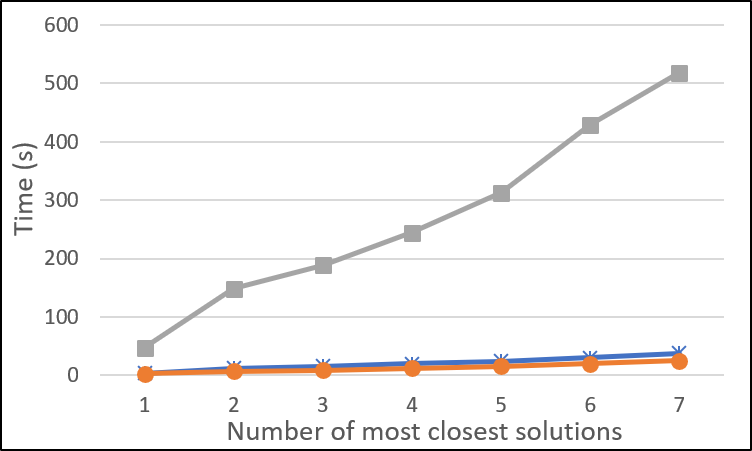}
		\caption{Comparison on performance.}
	\end{subfigure}
	\caption{Compare different $\alpha$-conversion techniques based on the same set of reference solutions.}
	\label{fig:aligns}
\end{figure*}

\subsection{In-depth Analysis}
We now present a few in-depth experiments to better understand the contributions of different techniques for the search, align, and repair phases. 
First, we investigate the effect of different program embeddings 
adopted in the search procedure. Then, we investigate the usefulness of our 
$\alpha$-conversion mechanism compared against other variable alignment approaches. For these experiments, we use two key metrics: 
(i) \emph{performance} (\ie how long does $\tool$ take to 
generate feedback), and (2) \emph{capability} (\ie how 
many incorrect programs are repaired with minimum 
fixes). To keep our experiments feasible, we 
focus on three problems that have the most 
reference solutions (\ie Printing, MaximumDifference 
and FibonacciSum). 

\paragraph{\textbf{Program Embeddings}} 
As the number $k$ of top-$k$ closest programs used for the reference
solutions increases, we compare the precision of our program
embeddings using position-aware characteristic vector against (1)
program embeddings using the characteristic vector and (2) using the
original AST representation.  We adopt a cut-off point of
$\mathcal{F}_{m}({P}_{e}, {P}_{\alpha\!c})$ to be three. Otherwise, if
we let $\tool$ traverse the power set of $\mathcal{F}({P}_{e},
{P}_{\alpha\!c})$, the capability criterion will be redundant. In
all settings, $\tool$ adopts the usage-set based $\alpha$-conversion
via position-aware characteristic vectors. Also, $\tool$ runs 
without any optimization techniques.  Figure~\ref{fig:vectors}
shows the results. The embeddings using position-aware characteristic
vectors are almost as precise as ASTs and achieves exactly the same
accuracy as ASTs when the number of closest solutions equals
five.  In addition, the position-aware characteristic vector embedding
consistently outperforms the characteristic vector embedding by more than
10\% in terms of capability. Moreover, although \pavector uses higher
dimensions to embed ASTs, it is generally faster due to the higher
accuracy in selected reference solutions, and in turn lowering the
computational cost spent on reducing $\mathcal{F}({P}_{e},
{P}_{\alpha\!c})$ to $\mathcal{F}_{m}({P}_{e}, {P}_{\alpha\!c})$.
When $\tool$ uses fewer reference solutions, the two embeddings do not
display noticeable differences since the speedup offered by the later is
insignificant. Both embedding schemes are significantly faster than
the AST representation.

\paragraph{\textbf{$\alpha$-Conversion}} 
We next evaluate the precision 
and efficiency of the $\alpha$-conversion in the align step. 
Given the same set of candidate solutions (identified by AST 
representation in the previous step), we compare our 
usage-based $\alpha$-conversion via position-aware 
characteristic vector against (1) the same usage-based 
$\alpha$-conversion via standard tree-edit operations and (2) a dynamic trace-based $\alpha$-conversion via sequence alignment. 
We adopt the same configurations as in the previous experiment 
which is to set the cut-off point of 
$\mathcal{F}_{m}({P}_{e}, {P}_{\alpha\!c})$ to be three and 
perform minimization without optimizations. As shown in 
Figure~\ref{fig:aligns}, our usage-set based $\alpha$-conversion via position-aware 
characteristic vectors outperforms that via standard tree-edit 
operations by more than an order of magnitude at the expense 
of little precision loss measured by capability. In the 
meanwhile, dynamic variable-trace based sequence alignment 
displays the best performance at a considerable cost of 
capability mainly due to its poor tolerance against erroneous 
variable traces.

\paragraph{\textbf{Optimizations}} 
One of the main performance bottlenecks in the repair phase is dynamic
execution, where the largest performance hit comes from the traversal
of power sets of $\mathcal{F}({P}_{e}, {P}_{\alpha\!c})$.
To tackle this challenge, we design and realize two key techniques
that help prune the search space substantially. First, we leverage the
reachability of a fix to examine its applicability before it is
attempted. In particular, if a fix location is never reached on any
execution path \wrt any of the provided inputs, it is safe to exclude
it from the minimization procedure. Second, certain corrections are
coordinated, \ie they should only be included/excluded simultaneously
due to their co-occurrence nature.  According to our evaluation, these
optimizations are able to gain approximately one order of magnitude
speedup.

\subsection{Reliance on Data}
\begin{figure}[thb!]
	\begin{subfigure}[b]{0.23\textwidth}
		\includegraphics[width=\textwidth]{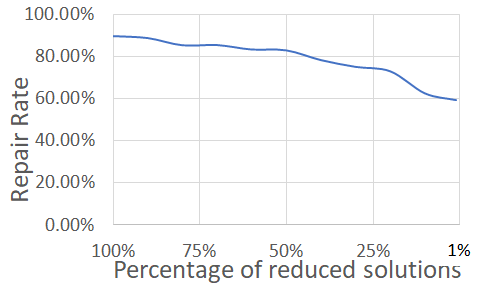}
		\caption{Capability trend.}
		\label{fig:capred}
	\end{subfigure}
	\begin{subfigure}[b]{0.23\textwidth}
		\includegraphics[width=\textwidth]{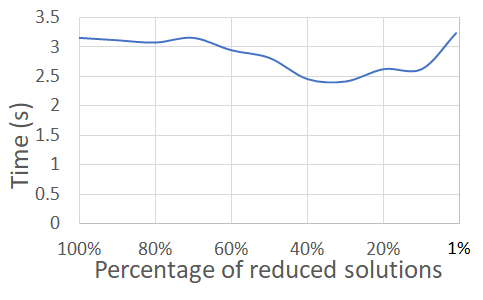}
		\caption{Performance trend.}
		\label{fig:perred}
	\end{subfigure}
	\caption{Capability and performance changes as the number of correct solutions decrease.}
	\label{fig:red}
\end{figure}

We conducted a further experiment to understand 
the degree to which $\tool$ relies on the correct 
programming submissions to have a reasonable 
utility. Initially, we use all the correct programs 
from all the programming problems, then we gradually 
down-sample them to observe the effects this may have 
on $\tool$'s capability and performance. 
Figure~\ref{fig:capred}/\ref{fig:perred} depicts the capability/performance 
change as the number of correct solutions is reduced 
from 100\% to 1\% under the standard configuration. In terms of capability, $\tool$ maintains 
almost the same power as the number of correct programs 
drops to half of the total. Even using only 1\% of the 
total correct programs, $\tool$ still manages to produce 
feedback based on minimal fixes for almost 60\% of 
the incorrect programs in total. 
The reason for this phenomenon is that the vast majority 
of students generally adopt an algorithm that their 
peers have correctly adopted. So even though the correct 
programs are down-sampled, there 
still exist some solutions of common patterns that can help 
a large portion of students who also attempt to 
solve the problem in a common way. Consequently, 
those students will not be affected. On the other 
hand, $\tool$ will understandably have more difficulties 
to deal with corner-case programming submissions. However, 
because the number of such programs is small, it 
generally does not have a severe impact. As 
for performance, the changes are twofold. 
On one hand, with fewer correct solutions, 
$\tool$ performs less computation. On the other hand,
$\tool$ generally spends more time reducing 
$\mathcal{F}({P}_{e}, {P}_{\alpha\!c})$ 
to
$\mathcal{F}_{m}({P}_{e}, {P}_{\alpha\!c})$ 
since the reference solutions become 
more dissimilar due to down-sampling. 
Since $\tool$ is able to find precise reference 
solutions for most of the incorrect programs when 
the number of reference solutions is not too low, \ie 
above 1\%, the final outcome is that $\tool$ 
becomes slightly faster.

\begin{figure}[thb!]
	\begin{subfigure}[b]{0.23\textwidth}
		\includegraphics[width=\textwidth]{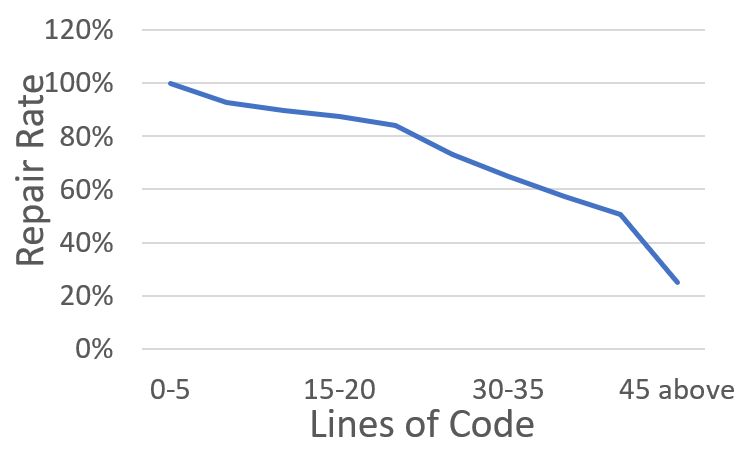}
		\caption{Capability trend.}
		\label{fig:capsca}
	\end{subfigure}
	\; 
	\begin{subfigure}[b]{0.23\textwidth}
		\includegraphics[width=\textwidth]{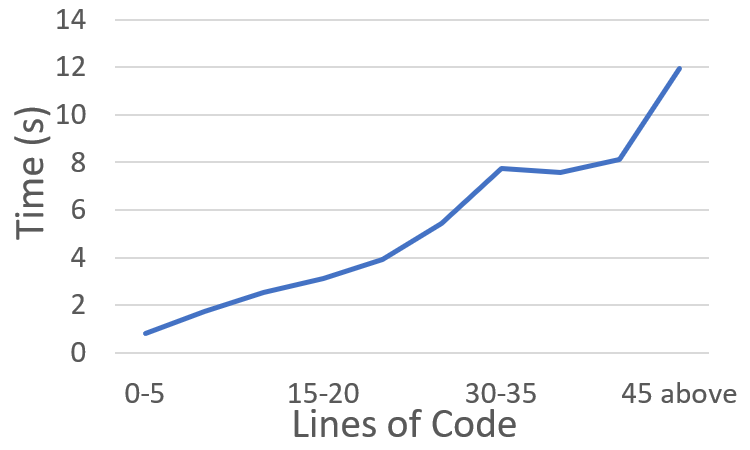}
		\caption{Performance trend.}
		\label{fig:persca}
	\end{subfigure}
	\caption{Capability and performance changes as the size of program grows.}
	\label{fig:sca}
\end{figure}

\subsection{Scalability with Program Size}
To understand how our system scales as the size 
of program increases, we conducted another experiment 
in which we partitioned the incorrect programs into 
ten groups according to size of the program, \ie the 
number of lines in code. In terms of performance (Figure~\ref{fig:capsca}), $\tool$ 
manages to generate the repairs within ten seconds in 
almost all cases. Regarding capability (Figure~\ref{fig:persca}), $\tool$ 
generates minimal fixes for close to 90\% of incorrect 
programs when their size is under 20 lines of code and 
still over 50\% when the size is less than 
40 lines of code. 

\subsection{Comparison against CLARA}
In this experiment, we conduct an empirical comparison against
CLARA~\cite{gulwani2016automated} using the same benchmark set. Since
CLARA works on C programs, we followed the following procedure.
First, we convert our \Csharp programs into C programs that CLARA
supports. In fact, we only converted the $\mathit{Console}$ operation
into $\mathit{printf}$ for the Printing problem, as a result we have
488 out of 742 programs from edX that still compile, but only 395 out
of 4,311 programs from CodeHunt since they generally contain more
complex data structures. In total, we have 883 programs as the
benchmark set for both systems to compare. Both systems use exactly
the same set of correct programs in different languages.  Second,
because we experienced issues when invoking the provided clustering
API\footnote{The match command provided in the Example section at
  https://github.com/iradicek/clara produces unsound result} to
cluster correct programs, we instead run CLARA on each correct program
separately to repair the incorrect programs and select the fixes that
are minimal and fastest (prioritize minimality over performance when
necessary). On the other hand, $\tool$ is set up with standard
configuration following Algorithm~\ref{alg:fg}.

The results are shown in Table~\ref{Table:com}. As the solution set is
down-sampled from 100\% to 1\%, $\tool$ generates consistently more
minimal fixes than CLARA (Table~\ref{Table:cap}). For the results on
performance shown in Table~\ref{Table:per}, CLARA outperforms $\tool$
marginally on the programs of small size, \ie, fewer than 15 lines of
code, whereas CLARA scales significantly worse than $\tool$ when the
size of programs grows, \ie, slower by more than one order of magnitude
when dealing with program of more than 25 lines. Regarding the
performance comparison, in reality, CLARA usually compares an
incorrect program with hundreds of reference solutions according
to~\cite{gulwani2016automated} to pick the smaller fixes, therefore
the performance measured is a significant
under-estimation. Furthermore, CLARA shows better performance in
part due to the less work it undertakes since it does not
guarantee minimal repairs.

\begin{table}[!htb]
	\begin{subtable}{.99\linewidth}
		\centering
 		\begin{adjustbox}{max width=\linewidth}
		\begin{tabular}{|c | c | c |} 
			\hline
			\begin{tabular}{@{}l@{}}\textbf{Percentage of} \\ \textbf{Solutions}\end{tabular} & 
			\begin{tabular}{@{}l@{}}\textbf{Number of Minimum} \\ \textbf{Fixes ($\tool$)}\end{tabular} & 
			\begin{tabular}{@{}l@{}}\textbf{Number of Minimum} \\ \textbf{Fixes (CLARA)}\end{tabular} \\
			\hline
			100\% &688 & 573 \\			
			\hline
			80\% &661 & 540 \\			
			\hline
			60\% &647 & 492 \\			
			\hline
			40\% &603 & 426 \\		
			\hline			
			20\% &539 & 349 \\			
			\hline
			1\% &466 & 271 \\
			\hline
		\end{tabular}
		\end{adjustbox} 			
	\caption{Comparison on capability.}
	\label{Table:cap}	
	\end{subtable}%
	\hfill	
	\begin{subtable}{.99\linewidth}
	\centering
	\begin{adjustbox}{max width=\linewidth}
		\begin{tabular}{|c | c | c |} 
			\hline
			\begin{tabular}{@{}l@{}}\textbf{LOC (Total \# of} \\ \textbf{Programs)}\end{tabular} &  
			\begin{tabular}{@{}c@{}}\textbf{Average Time Taken} \\ \textbf{($\tool$)}\end{tabular} & 
			\begin{tabular}{@{}c@{}}\textbf{Average Time Taken} \\ \textbf{(CLARA)}\end{tabular} \\
			\hline
			0-5 (122) & 1.14s & 0.84s \\
			\hline 								
			5-10 (227) & 1.38s & 1.71s \\
			\hline 								
			10-15 (269)& 2.69s & 2.25s \\
			\hline 								
			15-20 (137)& 3.02s & 9.71s \\
			\hline 								
			20-25 (91) & 3.81s & 20.84s \\
			\hline 								
			25 or more (37) & 5.26s & 40.39s \\
			\hline
		\end{tabular}
	\end{adjustbox} 			
	\caption{Comparison on performance.}
	\label{Table:per}	
	\end{subtable}%
	\caption{$\tool$ vs. CLARA on the same dataset.}
	\label{Table:com}	
\end{table}

\section{Related Work}
\label{sec:rel}

This section describes several strands of related work from the areas
of automated feedback generation, automated program repair, fault
localization and automated debugging.

\subsection{Automated Feedback Generation}
Recent years have seen the emergence of automated feedback generation
for programming assignments as a new, active research topic. We
briefly review the recent techniques.

\vspace{4pt}
\paragraph{\textbf{AutoGrader}}~\cite{singh2013automated}
proposed a program synthesis based automated feedback generation for
programming exercises. The idea is to take a reference solution and an
error model consisting of potential corrections to errors student
might make, and search for the minimum number of corrections using a
SAT-based program synthesis technique. In contrast, 
$\tool$ advances the technology in the following aspects: (1) $\tool$
completely eliminates the manual effort involved in the process of
constructing the error model; and (2) $\tool$ can perform more complex
program repairs such as adding, deleting, swapping statements, \etc

\vspace{4pt}
\paragraph{\textbf{CLARA}}~\cite{gulwani2016automated} 
is arguably the most similar work to ours. Their approach is to
cluster the correct programs and select a canonical program from each
cluster to form the reference solution set. Given an incorrect student
solution, CLARA runs a trace-based repair procedure \wrt each program
in the solution set, and then selects the fix consisting of the
minimum changes. Despite the seeming similarity, $\tool$ is
fundamentally different from CLARA. At a conceptual level, CLARA
assumes for every incorrect student program, there is a correct
program whose execution traces/internal states only differs because of
the presence of the error. Even though the program repairer generally
enjoys the luxury of abundant data in this setting, there are a
considerable amount of incorrect programs which yield new (partially)
correct execution traces.  Since the trace-based repair procedure does
not distinguish a benign difference from a fix, it will introduce
semantic redundancies which likely will have a negative impact on
student's learning experience. As we have presented in our evaluation,
CLARA scales poorly with increasing program size, and does not
generate minimal repairs on the benchmark programs.

\vspace{4pt}
\paragraph{\textbf{sk\_p}}~\cite{pu2016sk_p} was recently proposed
to use deep learning techniques for program repair. Inspired by the
skipgram model, a popular model used in natural language
processing~\cite{mikolov2013distributed, pennington2014glove}, sk\_p
treats a program as a collection of code fragments, consisting of a
pair of statements with a hole in the middle, and learns to generate
the statement based on the local context. Replacing the original
statement with the generated statement, one can infer the generated
statement contains the fix if the resulting program is
correct. However, sk\_p suffers from low capability results, as the
system only perform syntactic analysis. Another issue with the deep
learning based approaches is low reusability. Significant efforts are
needed to retrain new models to be applied across new problems.

\vspace{4pt}
\paragraph{\textbf{QLOSE}}~\cite{d2016qlose} is another recent work
for automatically repairing incorrect solutions to programming
assignments. The major contribution of this work is the idea of
measuring the program distance not only syntactically but also
semantically, \ie, preserving program behavior regardless of syntactic
changes. One way to achieve this is by monitoring runtime execution.
However, the repair changes to an incorrect program is based on a
pre-defined template corresponding to a linear combination of
constants and all program variables in scope at the program
location. As we have shown, more complex modifications are
necessary for real-world benchmarks.

\vspace{4pt}
\paragraph{\textbf{REFAZER}}~\cite{rolim2017learning} is 
another approach applicable in the domain of repairing program
assignments. The idea is to learn a syntactic transformation pattern
from examples of statement/expression instances before and after the
modification. Despite the impressive results, this approach also
suffers from similar issues as QLOSE, \ie, there are a large amount
of incorrect programs that require changes more complex than simple
syntactic changes.

\vspace{4pt}
\paragraph{\textbf{CoderAssist}}~\cite{Kaleeswaran} 
presents a new methodology for generating verified feedback for
student programming exercises. The approach is to first cluster the
student submissions according to their solution strategies and ask the
instructor to identify a correct submission in each cluster (or add
one if none exists). In the next phase, each submission in a cluster
is verified against the instructor-validated submission in the same
cluster.  Despite the benefit of generating verified feedback, there
are several weaknesses.  As mentioned, CoderAssist requires manual
effort from the instructor. More importantly, the quality of the
generated feedback relies on how similar the provided solution is to
the incorrect submissions in the same cluster.  In contrast, $\tool$
searches through all possible solutions automatically and uses those
that it considers to be the most similar to repair the incorrect
program.  In addition, CoderAssist targets dynamic programming
assignments only. Its utility and scalability would need require
further validation on other problems.

There is also work that addresses other learning aspects in the MOOC
setting. For example, Gulwani \etal~\cite{gulwani2014feedback}
proposed an approach to help students write more efficient algorithms
to solve a problem. Its goal is to teach students about the
performance aspects of a computing algorithm other than its functional
correctness. However, this approach only works with correct student
submissions, \ie, it cannot repair incorrect programs. Kim
\etal~\cite{Kim2016} is another interesting piece of work focusing on
explaining the root cause of a bug in students' programs by comparing
their execution traces. This approach works by first matching the
assignment statement symbolically and then propagating to match
predicates by aligning the control dependencies of the matched
assignment statements. The key difference is that our work can
automatically repair the student's code while Kim \etal~\cite{Kim2016}
can only illustrate the cause of a bug.


\subsection{Automated Program Repair}
Gopinath~\etal~\cite{Gopinath2011} propose a SAT-based approach to
generate repairs for buggy programs. The idea is to encode the
specification constraint on the buggy program into a SAT constraint,
whose solutions lead to fixes. K\"{o}nighofer and
Bloem~\cite{konighofer2011automated} present an approach based on
automated error localization and correction.  They localize faulty
components with model-based diagnosis and then produce corrections
based on SMT reasoning.  They only take into account the right hand
side (RHS) of the assignment statements as replaceable components.
Prophet~\cite{Long2016} learns a probabilistic,
application-independent model of correct code by generalizing a set of
successful human patches. There is also
work~\cite{Staber2005,Jobstmann2005} that models the problem of
program repair as a game. The two actors are the environment that
provides the inputs and a system that provides correct values for the
buggy expressions, so ultimately the specification is satisfied. These
approaches use simple corrections (\eg, correcting the RHS sides of
expressions) since they aim to repair large programs with arbitrary
errors.  Another line of approaches use program
mutation~\cite{debroy2010using}, or genetic
programming~\cite{Arcuri2008, Forrest2009} for automated program
repair. The idea is to repeatedly mutate statements ranked by their
suspiciousness until the program is fixed. In comparison our approach
is more efficient in pinpointing the error and fixes as those
mutation-based approaches face extremely large search space of mutants
($10^{12}$).

\subsection{Automated Debugging and Fault localization}
Test cases reduction techniques like Delta
Debugging~\cite{zeller2002simplifying} and
QuickXplain~\cite{Junker2004} can complement our approach by ranking
the likely fixes prior to dynamic analysis. The hope is to expedite
the minimization loop and ultimately speed up performance. A major
research direction of fault localization~\cite{Ball2003,
  groce2006error} is to compare faulty and successful executions. Jose
and Majumdar~\cite{Jose2011} propose an approach for error
localization from a MAX-SAT aspect. However, such approaches suffer
from their limited capability in producing fixes.

\section{Conclusion}
\label{section:conclu}

We have presented the ``Search, Align, and Repair'' data-driven
framework for generating feedback on introductory programming
assignments. It leverages the large number of available student
solutions to generate instant, minimal, and semantic fixes to
incorrect student submissions without any instructor effort. We
introduce a new program representation mechanism using position-aware
characteristic vectors that are able to capture rich structural
properties of the program AST. These program embeddings allow for
efficient algorithms for searching similar correct programs and
aligning two programs to compute syntactic discrepancies, which are
then used to compute a minimal set of fixes. We have implemented our
approach in the $\tool$ system and extensively evaluated it on
thousands of real student submissions. Our results show that $\tool$
is effective and improves existing systems \wrt automation,
capability, and scalability. Since $\tool$ is also language-agnostic,
we are actively instantiating the framework to support other languages such
as Python. $\tool$ has also been integrated with the
Microsoft-DEV204.1X edX course and the early feedback obtained from
online students demonstrates its practicality and usefulness.


\newpage

\bibliography{main}

%

\end{document}